\documentclass[11pt]{iopart}

\usepackage{graphicx}% Include figure files
\usepackage{dcolumn}% Align table columns on decimal point
\usepackage{color}
\usepackage{epstopdf}
\newcommand{\beq}{\begin{equation}}
\newcommand{\eeq}{\end{equation}}
\newcommand{\bea}{\begin{eqnarray}}
\newcommand{\eea}{\end{eqnarray}}
\newcommand{\p}{\partial}
\newcommand{\mb}{\mathbf}

\usepackage{iopams}  
\begin{document}

\title[Acceleration and Heating in a turbulent Corona]{Particle Acceleration and Heating in a Turbulent Solar Corona}

\author{Loukas Vlahos and Heinz Isliker}

\address{Department of Physics\\Aristotle University\\54124 Thessaloniki, Greece}
\ead{vlahos@astro.auth.gr,  isliker@astro.auth.gr}
\vspace{10pt}
\begin{indented}
\item[] June 2018
\end{indented}

\begin{abstract}
 Turbulence, magnetic reconnection, and shocks can be present in explosively unstable plasmas, forming a new electromagnetic environment, which we call here   \textit{turbulent reconnection}, and where spontaneous formation of current sheets takes place. 
 %inside this turbulence. 
 We will show that the  heating and the acceleration of particles is the result of the synergy of stochastic (second order Fermi) and systematic  (first order Fermi) acceleration inside fully developed turbulence. The solar atmosphere is magnetically coupled to a turbulent driver (the convection zone), therefore the 
 %formation 
 appearance
 of turbulent reconnection in the solar atmosphere is externally driven. Turbulent reconnection, once it is established in the solar corona, drives the coronal heating and particle acceleration.
 \end{abstract}

%\maketitle
\maketitle
%\tableofcontents

\section{Introduction \label{Intro}}
In the 80's, the link between  the spontaneous formation of current sheets inside fully developed turbulence and the evolution of unstable current sheets to fully developed turbulence has been established with the use of 2D numerical simulations of the MHD equations \cite{Matthaeus86, Biskamp89}. Several recent reviews discuss the way turbulence can become the host of reconnecting current sheets and how  reconnecting current sheets can drive turbulence \cite{Matthaeus11, Cargill12, Lazarian12, Karimabadi13a, Karibabadi2013c}. The link between shocks and large amplitude magnetic disturbances and current sheets has also been analyzed \cite{Karimabadi2014}.

We will use the term ``turbulent reconnection" to denote an environment where large scale magnetic discontinuities of size $\delta\!B$, with $\delta\! B/B >1$, coexist with randomly distributed Unstable Current Sheets (UCS) \cite{Matthaeus86, Lazarian99}. The importance of turbulent reconnection in many space and astrophysical systems has been discussed in detail  in recent reviews \cite{Lazarian15, Matthaeus15}.

In the solar atmosphere, turbulence is externally driven by the convection zone, and the spontaneous formation of a turbulent reconnecting environment has been analyzed in several articles \cite{Parker83, Parker88, Galsgaard96, Galsgaard97a, Gasgaard97b, Einaudi94a, Georgoulis96}.

This review is divided into three sections. In the first section we pose the question: How the three well known non linear MHD structures appearing in many astrophysical and laboratory plasmas, i.e.\ Turbulence, Current Sheet(s), and  Shocks, can lead asymptotically to turbulent reconnection. We will outline  briefly the current literature,  which addresses this question with the use of MHD, Hybrid and Particle In Cell (PIC) simulations. In section 3 we analyze the question: How the solar convection zone drives fully developed turbulence in the solar atmosphere. Finally, in  section 4, we attempt to reply to the  question: How the plasma is heated and the high energy particles are accelerated by turbulent reconnection. In section 5, we summarize our main points.

%\part{Turbulent Reconnection}

\section{On turbulent reconnection} \label{turrec}
\subsection{From turbulence to reconnection}
In most astrophysics applications ``turbulence" is represented by a collection of plasma normal modes, described by their dispersion relation,  i.e.\ an ensemble of low amplitude waves with random phases is considered \cite{Miller97}.
 We will call this representation ``weak-turbulence''. Its role in most astrophysical settings is rather limited, since the MHD waves will grow till they reach large amplitudes during most of the phenomena that are explosive or strongly driven.
Isliker et al. \cite{Isliker17a}  consider a strongly turbulent environment as it naturally results from the nonlinear evolution of the MHD equations, in a similar approach as in Dmitruk et al. \cite{Dmitruk04}. Thus, they did not set up a specific geometry of a reconnection environment or prescribe a collection of waves \cite{Arzner04} as turbulence model, but allow the MHD equations themselves to build
naturally correlated field structures (which are turbulent, not random) and
coherent regions of intense current densities (current filaments or CS).

The 3D, resistive, compressible and normalized MHD equations Isliker et al. \cite{Isliker17a} used are
\beq
\p_t \rho = -\nabla \cdot \mathbf{p}
\eeq
\beq
\p_t \mathbf{p} =
- \mathbf{\nabla}  \cdot
\left( \mathbf{p} \mathbf{u} - \mathbf{B} \mathbf{B}\right)
-\nabla P - \nabla B^2/2
%+ \rho \bar{\nu} \nabla^2 \mb{v}
\eeq
\beq
\p_t \mathbf{B} =
-  \nabla \times \mathbf{E}
\eeq
%\beq
%\p_t \mathcal{E} = -\mathbf{\nabla} \cdot \left[\mathcal{E}\mathbf{u} +
%P\mathbf{u} +
%\mathbf{E}\times\mathbf{B}\right]
%\eeq
\beq
\p_t (S\rho) = -\mathbf{\nabla} \cdot \left[S\rho \mathbf{u}\right]
\eeq
with $\rho$ the  mass density, $\mathbf{p}$ the momentum density,
$\mathbf{u} = \mathbf{p}/\rho$,
$P$ the thermal pressure,
$\mathbf{B}$ the magnetic field,
\beq \mathbf{E}   = -  \mathbf{u}\times \mathbf{B} + \eta \mathbf{J}\eeq
the electric field,
$\mathbf{J} =  \mathbf{\nabla}\times\mathbf{B}$
the current density, $\eta$ the resistivity,
%$\mathcal{E} = P/(\Gamma -1) +\rho u^2/2 + B^2/2$ the total energy density,
$S=P/\rho^\Gamma$ the entropy,
and $\Gamma=5/3$ the adiabatic index.

Isliker et al. \cite{Isliker17a} solved the 3D MHD equations numerically
(with the pseudo-spectral method \cite{Boyd2001}, combined with the strong-stability-preserving Runge Kutta scheme \cite{Gottlirb98}) in Cartesian coordinates and by applying periodic boundary conditions to a grid of size $128\times 128\times 128$. As initial conditions they use a fluctuating magnetic field $\vec{b}$ that consist of a superposition of Alfv\'en waves, with a Kolmogorov type spectrum in Fourier space, together with
a constant background magnetic field $\vec B_0$ in the
$z$-direction, so the total magnetic field is $\vec{B}=\vec{B}_0 +\vec{b}(x,y,z,t)$.  The mean value of the initial magnetic perturbation is  $<b> = 0.6 B_0$, its standard deviation is $0.3 B_0$, and the maximum equals $2B_0$, so that they indeed consider strong turbulence. The initial velocity field is 0, and the initial pressure and energy are constant.

The structure of 
%the magnetic field and 
the $z$-component of the current density $J_z$ is shown in Fig.\ \ref{f1.png}.
\begin{figure}[h]
	\centering
	\includegraphics[width=0.55\columnwidth]{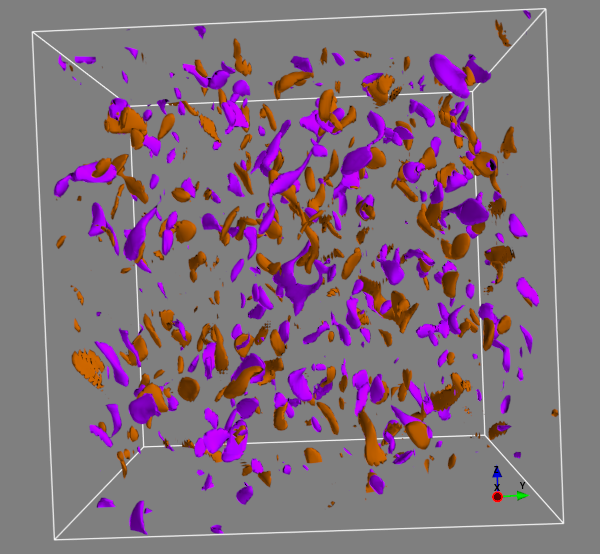}
	\caption{Iso-contours of the supercritical current density component $J_z$ (positive in brown negative in violet).  \cite{Isliker17a}. 
	%Bottom: Probability density function of the magnitude of the parallel
	% electric field.
	}
	\label{snapshot}
\end{figure}
For the MHD turbulent environment to build, Isliker et al \cite{Isliker17a} let the MHD equations evolve %for one Alvf\`en crossing time.
until the largest velocity component starts to exceed twice the Alvf\`en speed.
The magnetic Reynolds number at final time is $<|\mathbf{u}|> l /\eta = 3.5\times 10^3$, with $ l \approx 0.01$ a typical eddy size.
%, and the ratio of the energy carried by the magnetic perturbation to the
% kinetic energy is $(0.5 <b^2>) /(0.5<\rho \mathbf{u}^2>) = 1.4$, which is
% a second indication that we consider strong turbulence.
The overall picture in Fig.\ \ref{snapshot} demonstrates the spontaneous formation of current sheets. This result resembles the 2D simulations of Biskamp and Walter \cite{Biskamp89} almost thirty years ago. The perpendicular component of the current fluctuates rapidly but lacks the coherent structures shown in $J_z$. Similar results were obtained by Arzner et al. \cite{Arzner04,   Arzner06}, using Gaussian fields or the large eddy simulation scheme.

\begin{figure}[h]
	\centering
	\includegraphics[width=0.45\columnwidth]{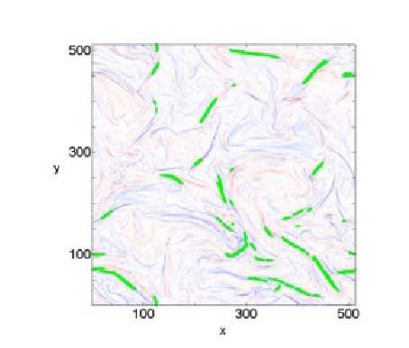}
	\includegraphics[width=0.50\columnwidth]{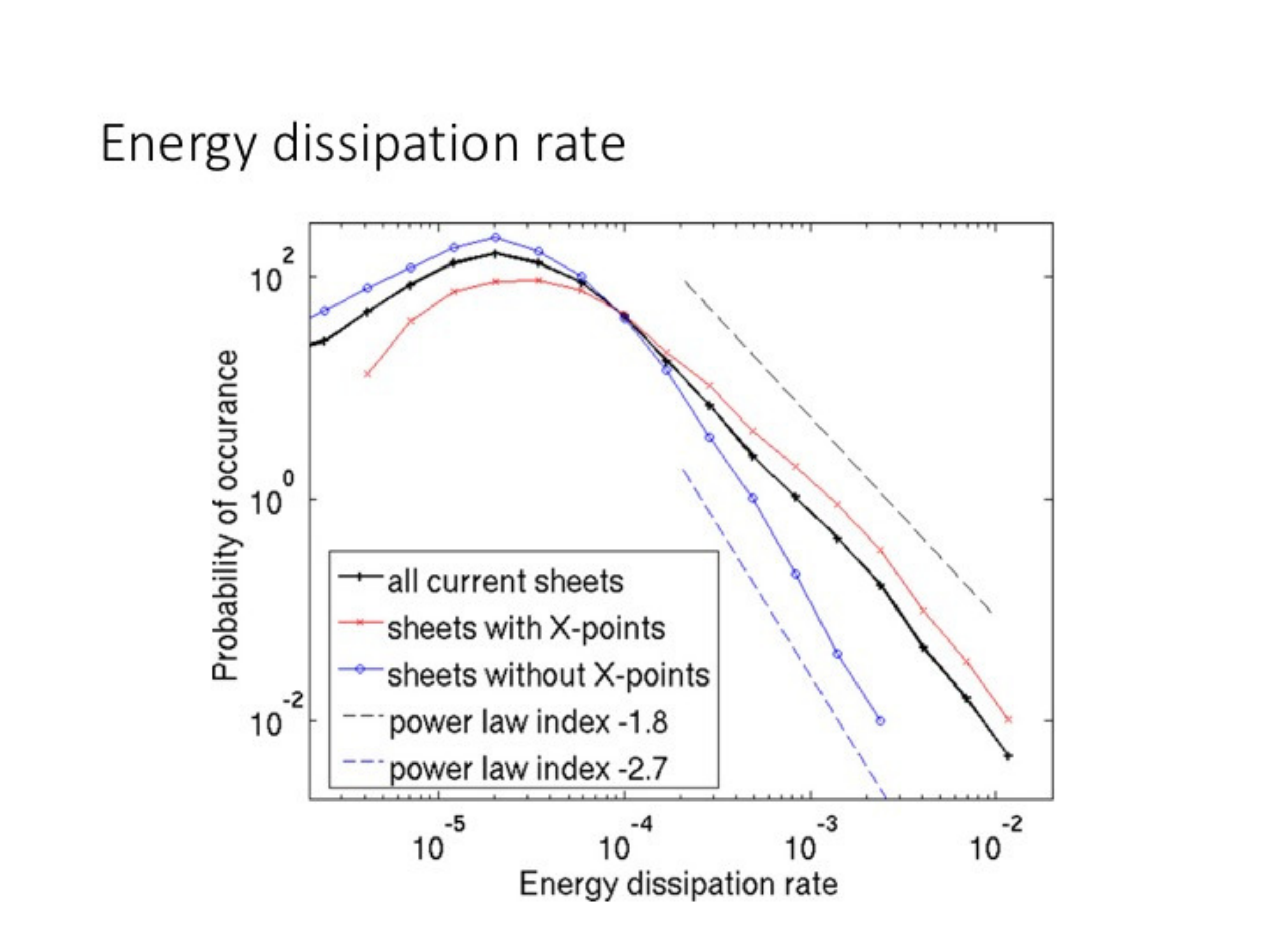}	
	\caption{Current density in an $x$-$y$-plane cross section of data. Red indicates negative current and blue indicates positive current. Identified current sheets in the plane are marked by green color.  (b) Probability distribution of the current sheet Ohmic dissipation rate. The distribution from all current sheets (black) shows a power law tail with index near $-1.8$. (From \cite{Zhdankin13}.) 
	%\cite{Isliker17a}.
	}
	\label{Distrcs}
\end{figure}

The statistical properties of the current sheets formed inside strongly turbulent environments 
%and their statistical characteristics 
have been analyzed in depth in 2D and 3D simulations by many researchers \cite{Servidio09, Servidio10,  Servidio11, Uritsky10,   Zhdankin13}. Zhdankin et al. \cite{Zhdankin13} developed a framework for studying the statistical properties of current sheets formed inside a magnetized plasma using a 3D reduced MHD code. The current fragmentation in an $x$-$y$-plane, which includes current sheets, is shown in Fig. \ref{Distrcs}. They were able to show that a large number of current sheets do not contain reconnection sites, and likewise, many reconnection sites do not reside inside current sheets.
The most striking characteristic of the current sheets formed spontaneously inside the strongly turbulent plasma is the probability distribution of the dissipated energy $\varepsilon =\int \eta j^2 dV$, which follows a power-law in shape, as reported by Zhdankin et al. \cite{Zhdankin13}
(see Fig.\ \ref{Distrcs}).

\subsection{From reconnection to turbulence}
\label{reconnection2turbulence}

The simple scenario for magnetic reconnection starts from a single reconnecting layer with all field lines smooth and well behaved. The conditions on both sides of the inflow, far from the reconnection zone, are assumed quiescent. This ``monolithic'' and highly idealized scenario for reconnection has been discussed extensively \cite{Priest14}.

Onofri et al. \cite{Onofri06} numerically solved the incompressible, dissipative, magnetohydrodynamics
(MHD) equations in dimensionless units
in a three-dimensional Cartesian domain,
with kinetic and magnetic Reynolds numbers $R_v=5000$ and $R_M=5000$.
They set up the initial condition in such a way as to have a plasma that is at rest,
in the frame of reference of the computational domain, permeated by a
background magnetic field sheared along the $\hat{x}$ direction,
with a current sheet in the middle of the simulation domain.
They perturb these equilibrium fields with three-dimensional  divergence-free
fluctuations.
\begin{figure}[h!]
\centering
\includegraphics[width=0.35\columnwidth]{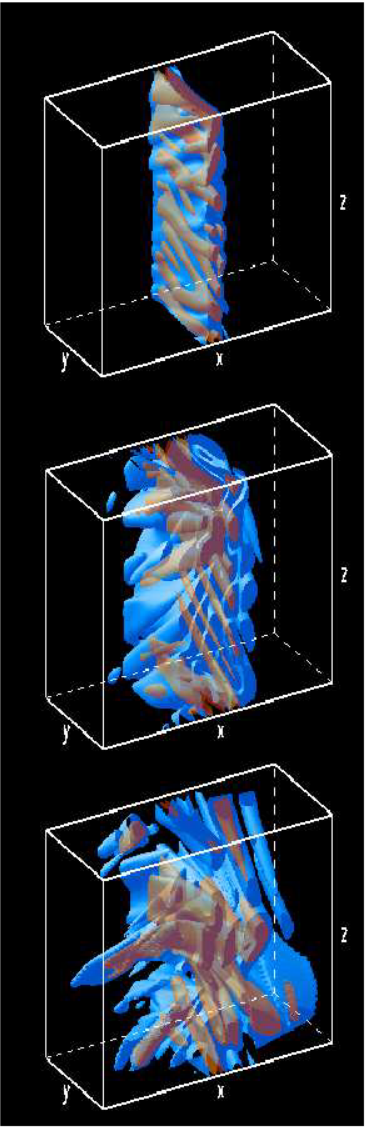}\caption{Electric field isosurfaces at $t=50\tau_A$,
$t=200\tau_A$ and $t=400\tau_A$. (From \cite{Onofri06}.)}
\label{figure1}
\end{figure}

The nonlinear evolution of the system is characterized by the formation
 of small scale structures, especially in the lateral
regions of the computational domain,
and by the coalescence of magnetic islands in the center.
This behavior is reflected in
the three-dimensional structure of the electric field,
which shows that
the initial equilibrium is destroyed by the formation of current filaments.
%For details about this MHD simulation see \cite{Onofri04}.
Figure \ref{figure1} shows the isosurfaces of the electric field at
different times calculated for two different values of the
electric field: the red surface represents higher values and the
blue surface represents lower values.
%The structure of the electric field is characterized by small regions of space
%where the  field is stronger, surrounded by a larger volume occupied
%by lower electric field values.
After about $t=50 \tau_A$ (where $\tau_A$ is the Alfv\'en time),
the current sheet starts to be fragmented.
%, as can be seen
%in Fig.\ \ref{figure1}, where we show the configuration of the electric field
%${\bf E}=\eta{\bf J}-{\bf v}\times{\bf B}$ calculated from the MHD simulation.
At later times the fragmentation is more evident,
and at $t=400\tau_A$,  the initial current sheet has been
completely destroyed and the
electric field is highly fragmented as well. Also Dahlin et al. \cite{Dahlin15}, by using kinetic simulations of a 3D collisionless plasma with a guide field, analyze the fragmentation of current sheets and the formation of small scale filaments with strong electric fields.

The way a large scale reconnecting current sheet(s)  evolves to reach the stage of turbulent reconnection when co-existing with turbulence has been analyzed by several authors. Among others a prominent mechanism identified is the appearance of multiple reconnection sites inside the turbulent region \cite{Matthaeus86, Lazarian99,  Drake10, Hoshino12b, Burgess16}.

It seems clear now, since the presence of wave activity leads to the collapse of a monolithic current sheet, as it can be seen in Fig.\  \ref{figure1} and the review of Ref.\ \cite{Cargill12}, that the simple assumptions of the monolithic reconnecting current sheet \cite{Priest14}, based on laminar and steady converging flows, may be  an unrealistic approximation for the energy release and particle acceleration in many natural or laboratory circumstances of dynamic plasmas.

\subsection{From shocks to turbulent reconnection}
The role of turbulent reconnection in shocks has not been analyzed in detail. Preliminary studies though suggest that there is a link between strong turbulence, reconnection and shocks \cite{Karimabadi2014,Matsumoto15}. The presence of ``turbulence'' as the source of converging scatterers upstream and downstream of a quasi parallel shock has been used extensively for the analysis of the Diffusive Shock Acceleration (DSA). But once again, such a monolithic isolated shock with externally prescribed turbulent flows \cite{Drury83} can not describe most of the realistic situations in astrophysical and laboratory plasmas.

The growth of unstable waves driven by accelerated ions \cite{Bell78a}, or pre-existing large amplitude fluctuations upstream of a shock, as e.g.\ in the solar wind flowing through the Earth's Bow shock,
drive turbulent reconnection upstream and especially downstream of the shock.

Numerical simulations cannot uncover the complexity of large scale shocks, nor can they follow its evolution  for long times \cite{Caprioli14a,Caprioli14b,Caprioli14c}, but the ignition of a turbulent reconnection environment has become clear  (see \cite{Karimabadi2014, Zank15, Matsumoto15} and the references listed in these articles). Matsumoto et al.\ \cite{Matsumoto15}, using  PIC  simulations, showed that strong collisionless shocks drive turbulent reconnection downstream of the shock.

\subsection{Turbulent reconnection as a state of strongly turbulent plasma}

In this section, we presented evidence  which suggests that three well know non linear plasma systems, (1) strong turbulence, (2) reconnecting current sheet(s) in the presence of waves, and (3) shocks, will evolve asymptotically into large scale systems in the state of turbulent reconnection, where large amplitude MHD disturbances and current sheets co-exist.

Turbulent reconnection in many space and astrophysical systems is externally driven, e.g.\ in the solar atmosphere, the solar wind, the Earth's magnetosheath, the Earth's magnetotail, etc. In the next section, we discuss the magnetic coupling of the turbulent solar convection zone with the solar atmosphere.

%\part{Turbulent reconnection  inside the Solar Corona}
\section{Turbulent reconnection in the Solar Corona}

\subsection{Convection zone driven turbulent reconnection in the solar atmosphere}

\begin{figure}[h]
\centering
\includegraphics[width=0.45\columnwidth]{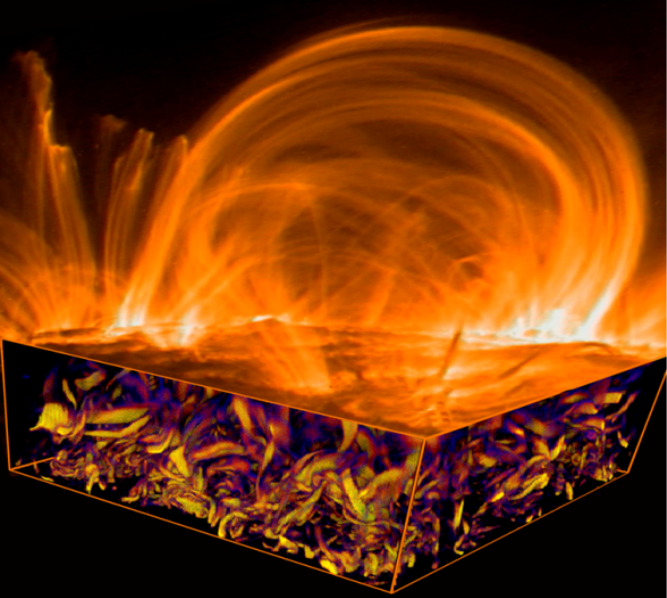}
\caption{ The magnetic link between the convection zone and the solar corona.}
\label{ar2}
\end{figure}

Parker \cite{Parker83}  was the first to realize that the spontaneous formation of magnetic current sheets in the solar corona is a  natural consequence of the magnetic link between the turbulent convection zone and the solar atmosphere \cite{Parker88, Parker94}.  In other words, the continuous random shuffling of the emerged magnetic field lines will drive reconnection (see Fig. \ref{ar2}a). In Fig. \ref{ar2}b we show a cartoon illustrating the link between the convection zone and the solar atmosphere.

Parker suggested in his original article that the spontaneous formation of current sheets in the solar corona by magnetic field braiding was part of the mechanism for the formation of the solar corona \cite{Parker83}. In 1983, Parker returned to his initial idea  that small scale current sheets are spontaneously formed inside the solar atmosphere, and used this concept as a cause for energy dissipation and source for coronal heating \cite{Parker88}. 

A few years after these ideas appeared, a series of articles analyzed, by using the MHD equations, the formation of current sheets, assuming a random forcing or motion of the magnetic field lines at the photosphere \cite{Mikic89,  Einaudi96,  Galsgaard96, Galsgaard97a, Gasgaard97b,  Georgoulis98, Rappazzo10}.

Einaudi et al. \cite{Einaudi96} analyzed a 2D section of a coronal loop, subject to random forcing of the magnetic fields. The title of their article was ``Energy release in a turbulent corona'', and they discussed the spontaneous formation of current sheets.  Geourgoulis et al. \cite{Georgoulis98} extended the simulations of Einaudi et al.\ for much longer times. Their aim was to extract reliable statistical information about turbulent reconnection in the solar atmosphere. Their main result was that the distribution function of both, the maximum and average current dissipation, and of the total energy content, the peak activity and the duration of such events, all show a robust scaling law, with scaling indices $\delta$ varying from $-1.9$ to $-2.8$ for temporal distribution functions, while $\delta \approx -2.6$ for spatial distributions of the dissipative events. Dmitruk and Gomez \cite{Dmitruk97} reached a similar conclusion using a two-dimensional lattice, with lower resolution.
% and obtain similar results.

\begin{figure}[h]
\centering
\includegraphics[width=0.85\columnwidth]{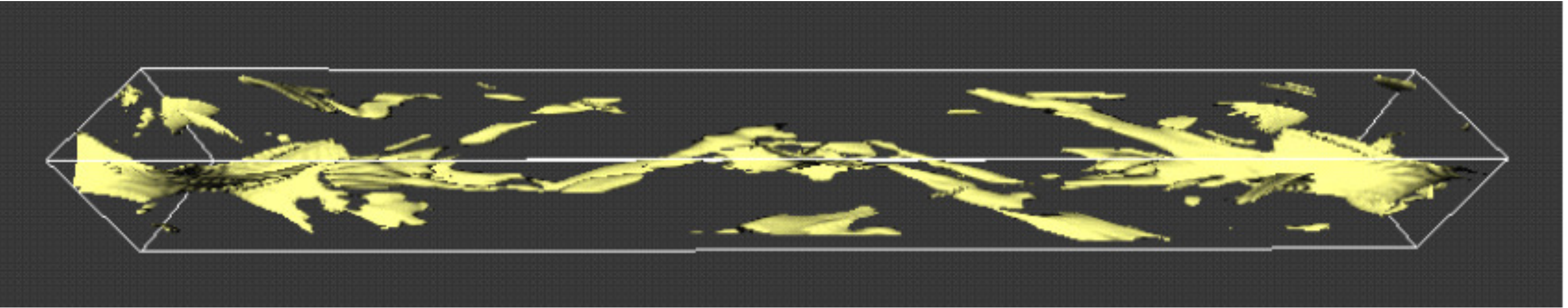}\\
\includegraphics[width=0.40\columnwidth]{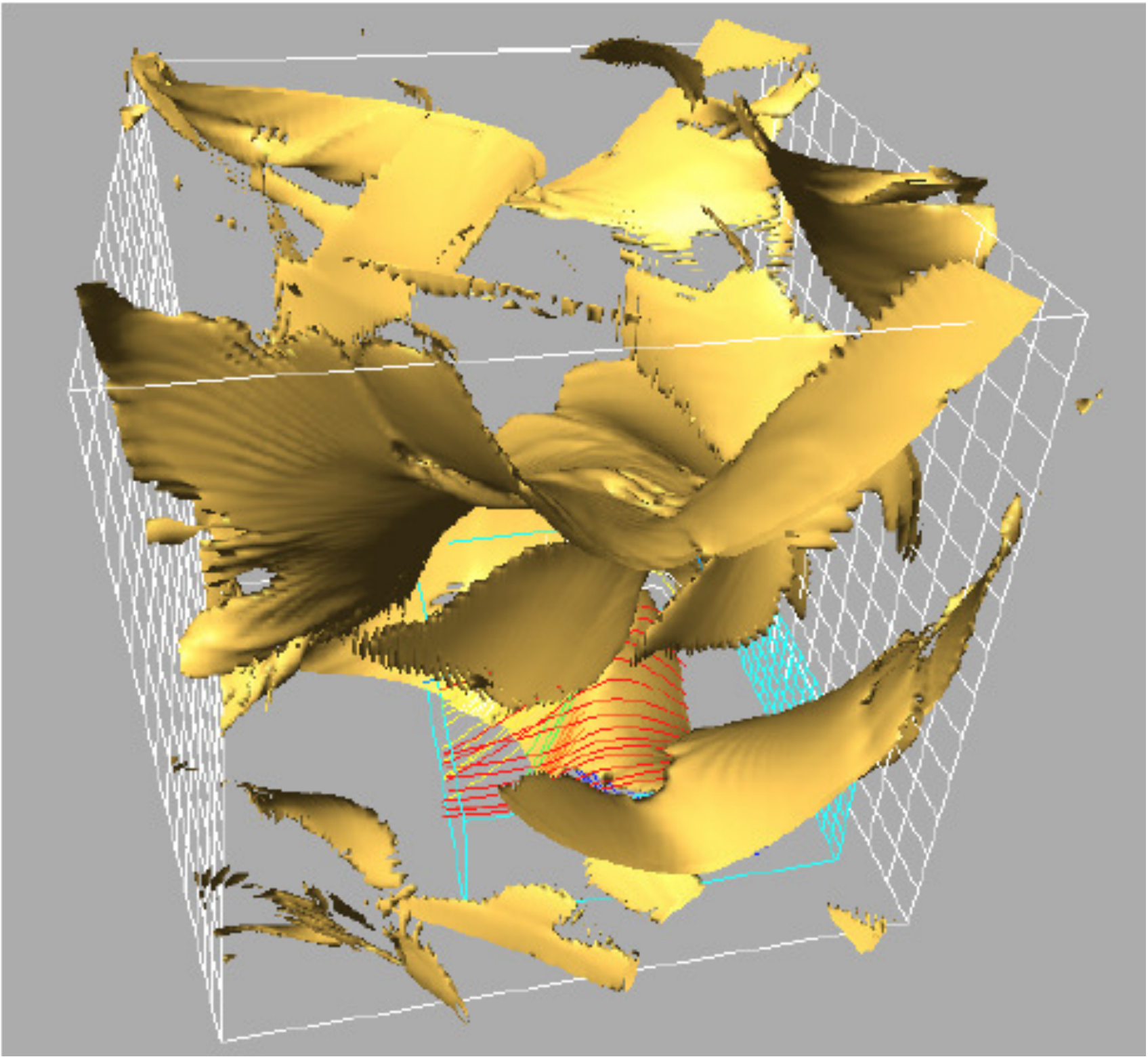}
\caption{(a) Isosurfaces of Joule dissipation regions, (b) isosurfaces of strong electric currents for a snapshot of the simulations. \cite{Galsgaard96}.}
\label{Dtr}
\end{figure}

Galsgaard and Nordlund \cite{Galsgaard96} solve the dissipative 3D MHD equations in order to investigate an initially homogeneous magnetic flux tube stressed by large scale sheared random motions at the two boundaries. The spontaneous formation of current sheets at random places and at random times inside the structure is shown in Fig.\ \ref{Dtr}.
Reconnection of the current sheet(s) straightens the field lines but also  causes large scale disturbances in the surrounding plasma, leading to further fragmentation of the energy release processes. Turbulent reconnection is established inside the magnetic flux tube and is driven by the random shear motions and the energy delivered to the solar corona through the dissipation of the fragmented current sheets (see \cite{Onofri06, Dahlin15}). The evolution of the turbulent reconnecting system depends on the velocities of the boundary motions and the initial magnetic field strength of the magnetic flux tube.

\begin{figure}[h]
\centering
\includegraphics[width=0.85\columnwidth]{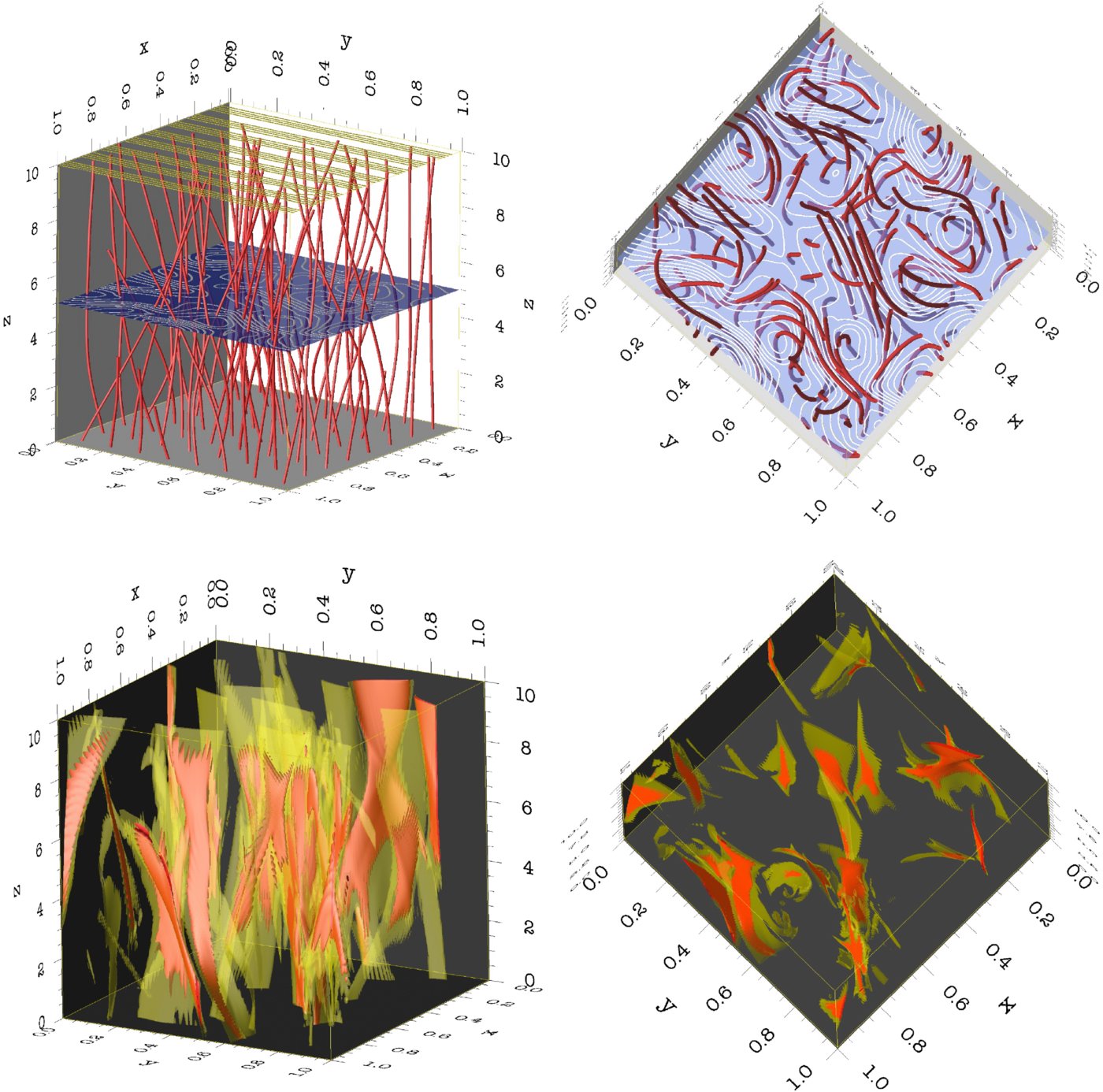}\\
\caption{Side and top views of a snapshot of magnetic field lines(top row) and current sheets (bottom row). See details in the original article on
%outline 
the origin of turbulence in the solar corona \cite{Rappazzo10}.}
\label{turb}
\end{figure}

In a series of articles, Rappazzo et al.\  and others \cite{Rappazzo10, Rappazzo13, RappazzoParker13, Velli15,  Rappazzo17, Warnecke17}, following the steps of the work of Garlsgaard and Nordlund \cite{Galsgaard96}, analyzed  the process of establishing  turbulent reconnection in the solar corona (see Fig.\ \ref{turb}). The observational expectations from the intermittent heating in turbulent reconnection were also investigated in depth \cite{Dahlburg16}.

\subsection{Turbulent reconnection during large scale reorganizations of the magnetic field in the solar atmosphere}

The development of a large scale MHD instability through the systematic driving of an emerged magnetic topology, or through the emergence of new magnetic flux from the convection zone, is another way turbulent reconnection can be driven in the solar atmosphere \cite{Hood79, Hood92, Torok04}. Gordovsky and Browning \cite{Gordovsky11} use  an initially isolated magnetic flux tube with uniform  magnetic field. The flux tube is stressed at both ends in different directions and becomes kink unstable. For our review here, the most important part of their analysis 
%, which is important 
is the topological evolution of the flux tube, since when the magnetic dissipation becomes significant and the connectivity between the two boundaries changes from ordered to chaotic, then turbulent reconnection has been established. In a similar study \cite{Rappazzo13}, a flux tube is stressed at the two ends by localized photospheric vortical motion, which twists the coronal field lines, and the current fragmentation reaches again the state of turbulent reconnection.

Magnetic flux emergence and the subsequent eruption was studied in many articles. Most of the numerical studies have stopped their analysis at the formation of a large scale current sheet through the interaction of an emerging flux tube with the ambient magnetic field of the solar  atmosphere. Archontis and Hood \cite{Archontis14} analyze the formation of standard jets driven by magnetic flux emergence and its interaction with the ambient magnetic field. They have pushed their study much beyond the formation of the initial current sheet, and the appearance of  current fragmentation is obvious in the snapshots following the formation of the jet. 

The observations supporting the claim that the solar corona is permanently in a turbulent state, as driven  by the high beta solar convection zone, are numerous and are beyond the scope of this review. We mention only the articles \cite{Granmer07,  Asgari14, Granmer15, Raouafi16, Kontar17} here, and also refer to the references cited therein for a more complete picture.

All pieces of information reported above are extremely useful for the analysis of the heating and acceleration of the coronal plasma during eruptions and large scale reorganizations of 
the magnetic field of active regions, being manifested as coronal mass ejections and large flares.

\section{Particle heating and acceleration by turbulent reconnection}
In the previous sections we explored the way turbulent reconnection is established in unstable plasmas and how the solar convection zone drives the solar corona into the state of turbulent reconnection. In the following sections, we analyze the way test particles are energized inside a large scale system of turbulently reconnecting plasma. The feedback of the energized test particles on the MHD equations is currently an open numerical problem. 

\subsection{Particle heating and acceleration in turbulent reconnecting coronal plasma, using the electromagnetic fields from MHD simulations}
Ambrosiano et al. \cite{Ambrosiano88} were the first to analyse the evolution of test particles inside turbulent reconnection, by using the electromagnetic fields derived from the simulations of Matthaeus and Lamkin \cite{Matthaeus86}, reported here in section 2. Many years later several researchers returned to this problem and followed the evolution of a distribution of particles inside a snapshot of 3D MHD simulations of a spectrum of MHD waves \cite{Dmitruk03,  Dmitruk04, Arzner06}. 

Isliker et al. \cite{Isliker17a} use the simulations reported in section 2 to explore the evolution of test particles inside a large scale turbulent reconnection environment. The test-particles are tracked in a fixed snapshot of the MHD evolution, 
and the particles are evolved for short times,  so Isliker et al.\  do not probe the scattering of particles off waves, but the interaction with electric fields.
In this particular numerical experiment, anomalous resistivity effects were also taken into account.  
Physical units are  introduced by using the parameters $L=10^5\,$m for the box-size, $v_A=2\times10^6\,$m/s
for the Alfv\'en speed, and $B_0=0.01\,$T for the background magnetic field.
%Isliker et al.  apply a cubic interpolation of the fields at the grid-points to the actual particle positions.

%\paragraph{Test-particle simulations.}
The relativistic guiding center equations (without collisions) are used
for the evolution of the position $\mb{r}$ and  the parallel component $u_{||}$ of the relativistic 4-velocity of the particles  
The test-particles considered throughout are electrons.
Initially, all particles are located at random positions, and they obey a 
Maxwellian distribution 
$n(W, t=0)$ 
with temperature $T=100\,$eV. The simulation box is open, the particles can escape from it when they reach any of its boundaries.
\begin{figure}[h]
\centering
\includegraphics[width=0.40\columnwidth]{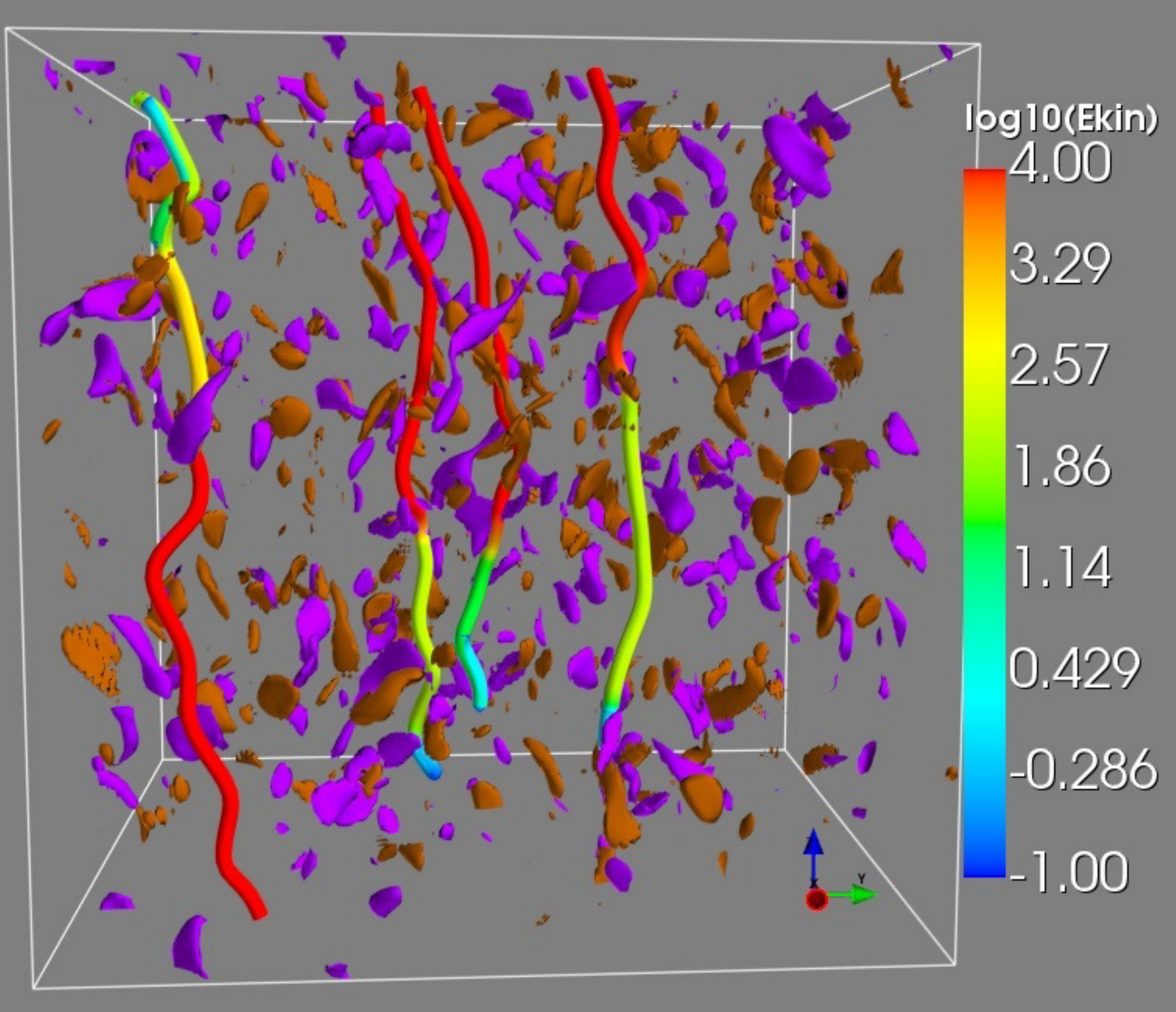}
\includegraphics[width=0.50\columnwidth]{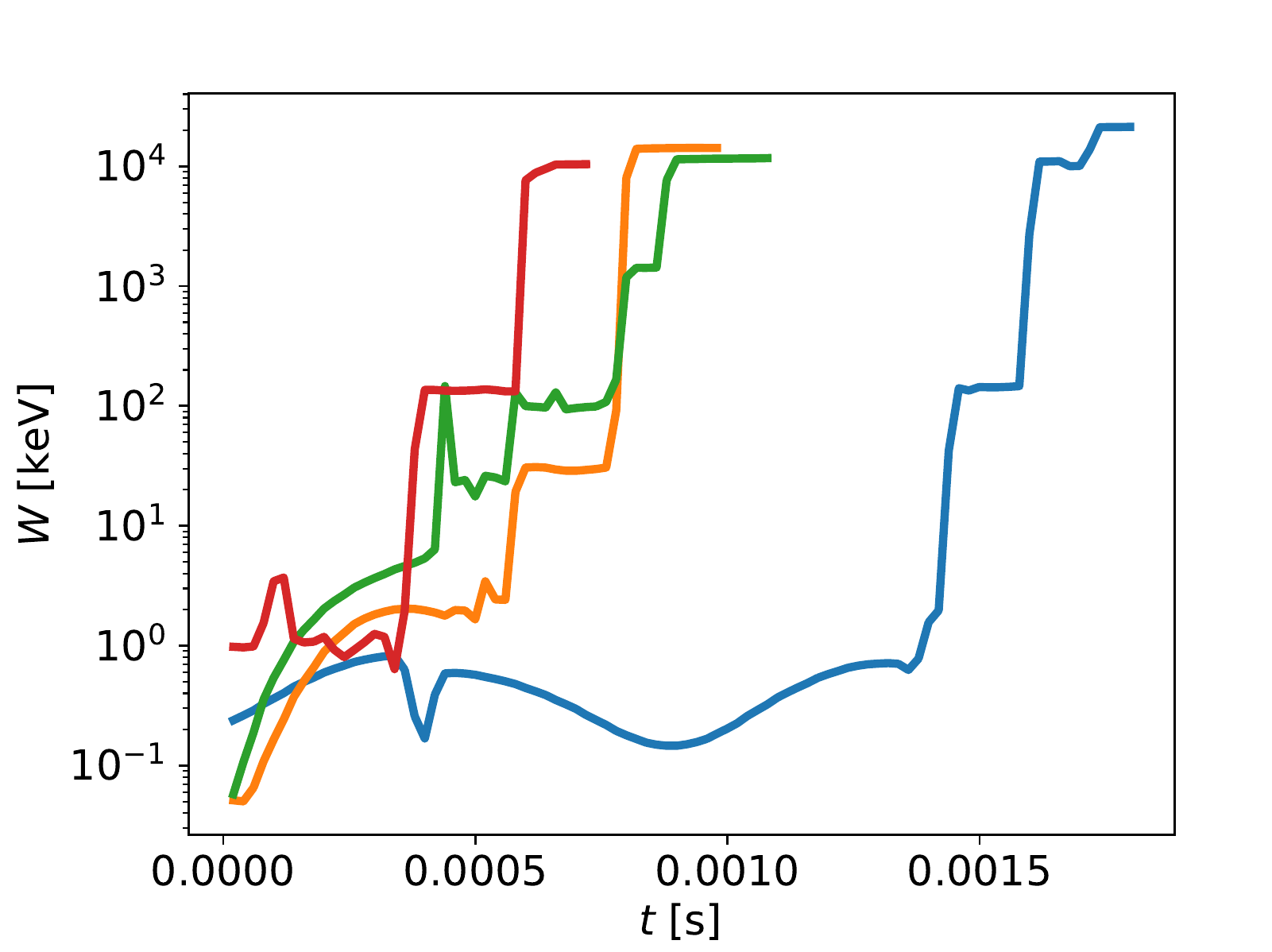}\\
\includegraphics[width=0.80\columnwidth]{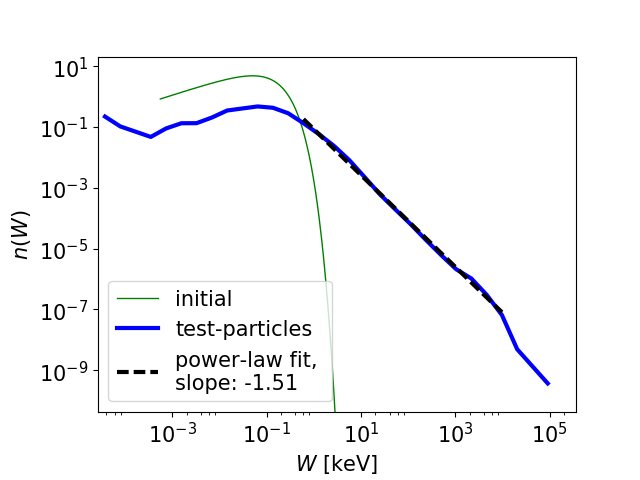}
\caption{ (a) Some particle orbits inside the simulation box, colored according to the logarithm of their kinetic energy. (b) Particle trajectories in energy space of the same energetic particles as in (a).  (c) Initial and final (at $t
			= 0.002\,$sec) kinetic energy distribution from the test-particle simulations, together with a power-law fit. \cite{Isliker17a}.}
\label{Ekin}
\end{figure}

The acceleration process, is very efficient, and Isliker et al. \cite{Isliker17a}  consider a final time
of $0.002\,$s, 
%($7\times 10^5$ gyration periods), 
at which the asymptotic state has already been reached.
As Fig.\ \ref{snapshot}, Fig.\ \ref{Ekin}a shows the component $J_z$ in the regions of above-critical current density, which clearly are fragmented into a large number of small-scale 
current filaments (current-sheets) that represent 
coherent structures within the nonlinear, super-Alfv\'enic MHD environment. 
%, up to the final time, 
%$t=0.8\,$s considered here, 
%or until they 
%escape from the simulation box. 

Fig.\ \ref{Ekin}a also shows a few orbits of energetic electrons inside the simulation box.
The particles can lose energy, yet they mostly gain energy in a number of sudden jumps in energy (see Fig.~\ref{Ekin}b), the energization process thus is localized and there is multiple energization at different current filaments. The acceleration thus is systematic and the particles
undergo a rapid increase of their energy most times when they pass through 
an UCS.
Fig.~(\ref{Ekin}c) shows the energy distribution at final time. It
exhibits a clear power law part in the intermediate to high
energy range with power-law index $-1.51$, with a slight turnover at the highest energies. 
There is also moderate heating, the initial temperature has roughly been
doubled
(qualitatively similar characteristics of the acceleration process
have been observed in \cite{Arzner04} and in the PIC simulations of
 \cite{Dahlin15,Guo15}.

Onofri et al.\ \cite{Onofri06} follow the evolution of test particles inside the electromagnetic fields from a 3D MHD simulation of a collapsing current sheet (see Sect.\ \ref{reconnection2turbulence} and Fig.\  \ref{figure1}). We already mentioned that the electric field is fragmented and the UCS asymptotically lead to a turbulent reconnection environment. To give a measure of the fragmentation of the electric field, Onofri et al.\  calculated the fractal dimensions of the fields shown in Fig.\ \ref{figure1},
using the box counting algorithm.

\subsection{Particle heating and acceleration by stochastic and systematic Fermi processes}

Fermi \cite{Fermi49,Fermi54} introduced two acceleration mechanisms for the heating and acceleration of plasmas, (1) stochastic acceleration (second order Fermi) by randomly moving "magnetic clouds", and (2) systematic acceleration (first order) by converging turbulent flows in the vicinity of a shock \cite{Longair11}.

The energy gain from the stochastic interaction of particles with  scatterers (magnetic clouds) is
\beq
\frac{\Delta W}{W} \approx \frac{2}{c^2}
(V^2-\vec{V}\cdot \vec{u})
\eeq
where $\vec{V}$ is the velocity of the scatterer, $\vec{u}$  the velocity of the test particle and c the speed of light.  Particles gain energy when $\vec{V}\cdot \vec{u}<0$ and loose energy when $\vec{V}\cdot \vec{u}>0. $ The rate particles gain energy is estimated from the relation 
$dW/dt \approx W/t_{acc}$, with $t_{acc} \approx (3 \lambda c)/(4 V^2)$, where $\lambda$ is the mean free path the particles travel between the scatterers.

A model to study the role of stochastic interaction of particles with large scale magnetic disturbances, as present in turbulent reconnection environments and analyzed in sections 2 an 3, was proposed in \cite{Vlahos16, Pisokas16}.

 \begin{figure}[h!]
\centering
\includegraphics[width=0.50\columnwidth]{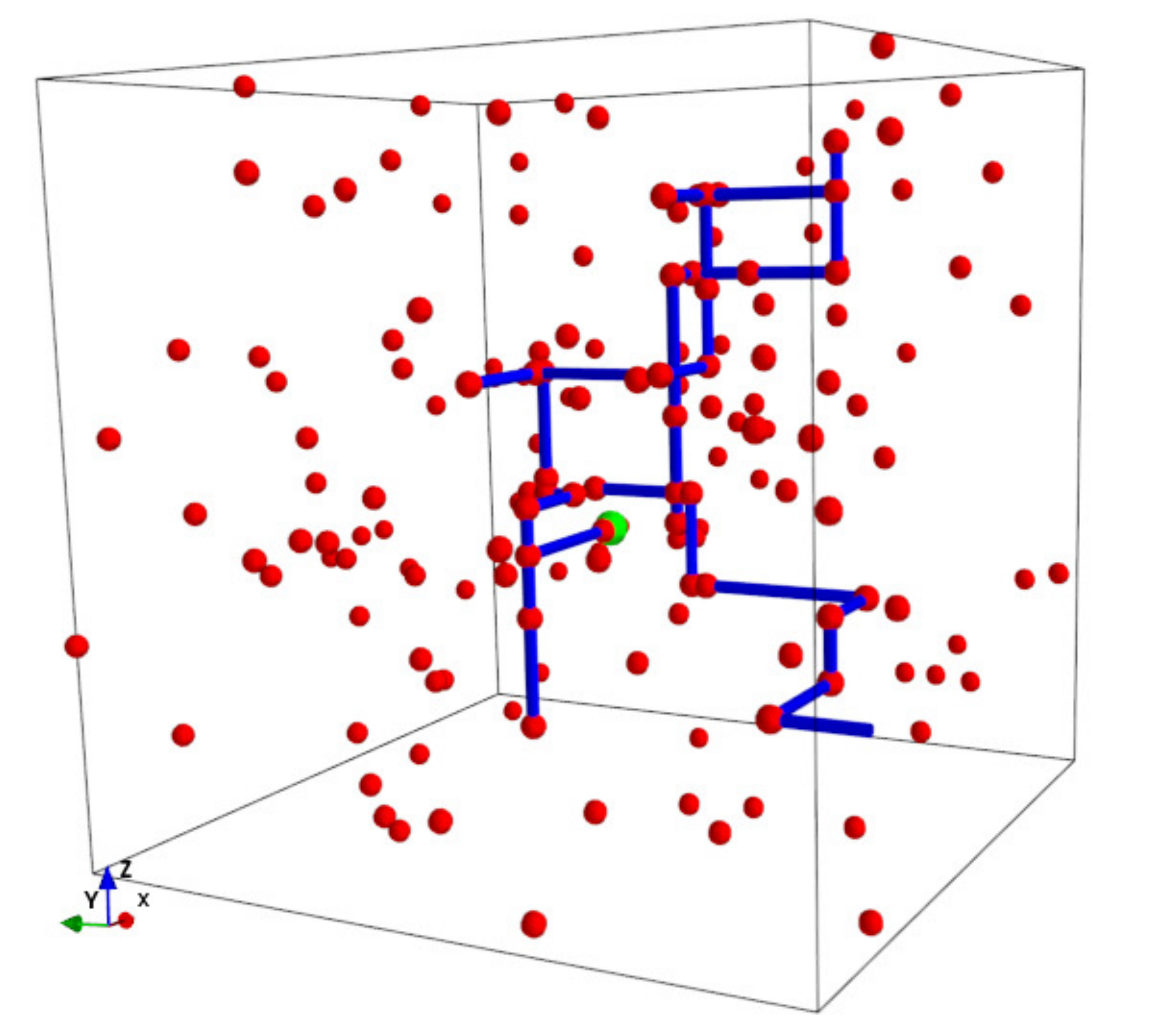}
\caption{The trajectory of a typical particle (blue tube) inside a grid with linear dimension $L$. Active points are marked by spheres in red color. The particle starts at a random grid-point (green sphere), moves along a straight path on the grid till it meets an active point, and then it moves into a new random direction, and so on, until it exits the simulation box. \cite{Pisokas16}.}
\label{Fermi1}
\end{figure}

Pisokas et al.\ \cite{Pisokas16}   construct a 3D grid $(N \times N \times N)$ with linear size $L$, with grid width $\ell=L/(N-1)$. Each grid point is set as either {\bf active} or \emph{inactive}, i.e.~a scatterer or not. Only a small fraction $R = N_{\rm sc}/N^3$ of the grid points are active (5-15{\%). The density of the scatterers can be defined as $n_{\rm sc} = R \times N^3/L^3$, 
and the mean free path of the particles between scatterers can be determined as $\lambda_{\rm sc}=\ell/R.$
When a particle (an electron or an ion) encounters an active grid point, it renews its energy state depending on the physical characteristic of the scatterer. It then moves into a random direction with its renewed velocity $v$, until it meets another active point or exits the grid. The minimum distance between two scatterers is the grid width ($\ell$). The time between two consecutive scatterings is $\Delta t = s/v,$ where $s$ is the distance the particle travels, and it is an integer multiple of the minimum distance $\ell$ (see Fig. \ref{Fermi1}).
At time $t = 0$ all particles are located at random positions on the grid. The injected distribution $n(W, t=0)$ is a Maxwellian with temperature $T$. The initial direction of motion of every particle is selected randomly.
The parameters used are related to the plasma parameters in the low solar corona. The strength of the magnetic field is  $B = 0.01 T$, the density of the plasma $n_0 = 10^9 cm^{-3}$, and the ambient temperature around $100 eV$. The Alfv\'en speed is $V_A \simeq 7 \times 10^8 cm/sec$, so $V_A$ is comparable with the thermal speed of the electrons. The energy increment is of the order of $\Delta W/W  \approx (V_A/c)^2 \sim 10^{-4}$  and the length $L$ of the simulation box  is $10^{10}cm$. The grid is considered to be open, so particles can escape from the acceleration region when they reach any boundary of the grid, at $t = t_{\rm esc}$, which is different for each escaping particle. It is assumed that only $R = 10{\%}$ of the $N^3 = 601^3$ grid points are active.
 \begin{figure}[h!]
\centering
\includegraphics[width=0.45\columnwidth]{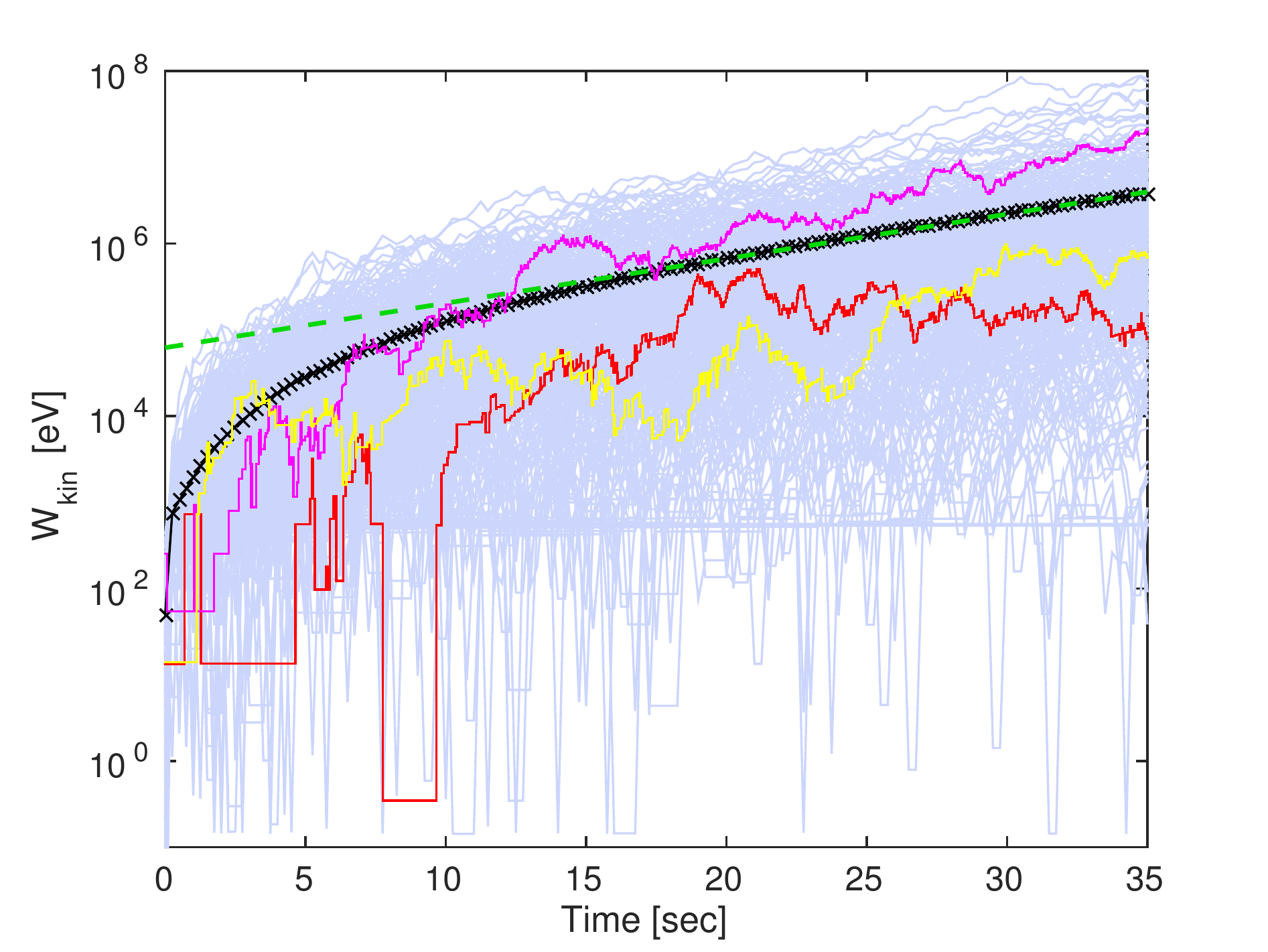}
\includegraphics[width=0.45\columnwidth]{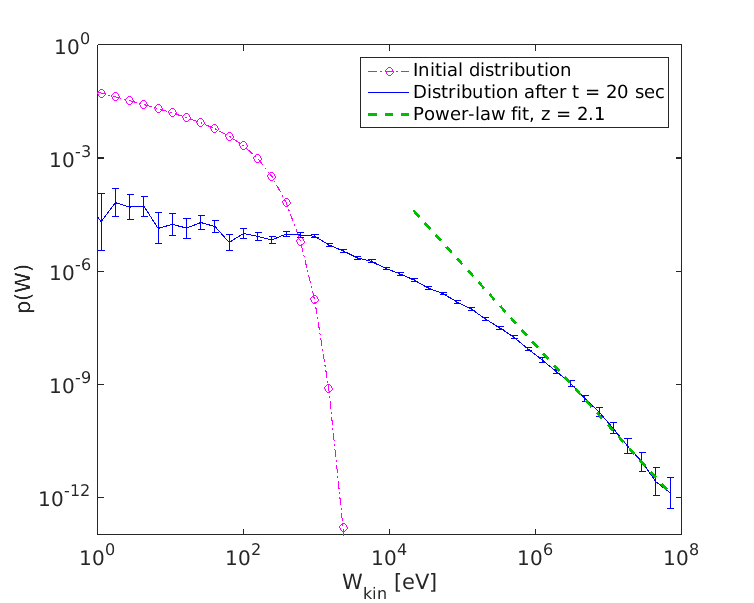}
\caption{ (a)  Kinetic energy of particles remaining inside the box as a function of time (blue) and their mean energy (black), with an exponential fit (green), and the kinetic energy of three typical electrons (colored). (b) Kinetic energy distribution at $t=0$ and $t=20\,$s (stabilised), with a power-law fit, for the electrons remaining inside the simulation box. \cite{Pisokas16}.}
\label{Fermi2}
\end{figure} 
The mean energy gain of the particles and the asymptotic energy distribution are shown in Fig.\ \ref{Fermi2}.  From the test particles, Pisokas et al.\ estimate $t_{acc}$ (see Fig.\ \ref{Fermi2}a) and the mean time $t_{esc}$ the particles remain inside the acceleration volume   (for details see \cite{Pisokas16}). It turns out that the results from this model agree very well with the predictions of the second order Fermi acceleration process, according to which the power law index of the energetic particles is 
\beq
k=1+\frac{t_{acc}}{t_{esc}} .
\eeq
For the parameters used, Pisokas et al.\ find $t_{acc} \sim  t_{esc}$, so that $k \sim 2$, which agrees well with the simulation results, see Fig.\ \ref{Fermi2}b}.

The key parameters controlling the heating and acceleration of particles in the described stochastic process are the density of the scatterers, the mean free path $\lambda$ of the particles traveling between the scatterers, the ambient magnetic field, and the size of the acceleration volume $L$. Using typical parameters from the solar corona, Pisokas et al.\  have shown  that stochastic interaction of particles with large amplitude magnetic fluctuations will also heat the plasma.   A parametric study of the evolution of the energy distribution of the particles, as we vary the density of the scatterers $0.05 < R < 0.15$ (i.e.~$1.1 \times 10^8 cm < \lambda_{\rm sc} < 3.3 \times 10 ^8 cm$), keeping the characteristic length of the acceleration volume constant, was made and we find that the escape time varies between $5 sec < t_{\rm esc} < 8 sec$, while the acceleration time decreases from $\simeq 8 sec$ to $\simeq 4sec$. The power-law tail index also decreases and it remains close to $3 \gtrapprox k \gtrapprox 1.5$. 
%In Fig.~\ref{f:F2o_R} we show stabilised distribution functions for $R=0.05$
% and $R=1.5$.  
The time when the $k$-index stabilizes varies between 20 and 25 sec.
 \begin{figure}[h!]
\centering
\includegraphics[width=0.45\columnwidth]{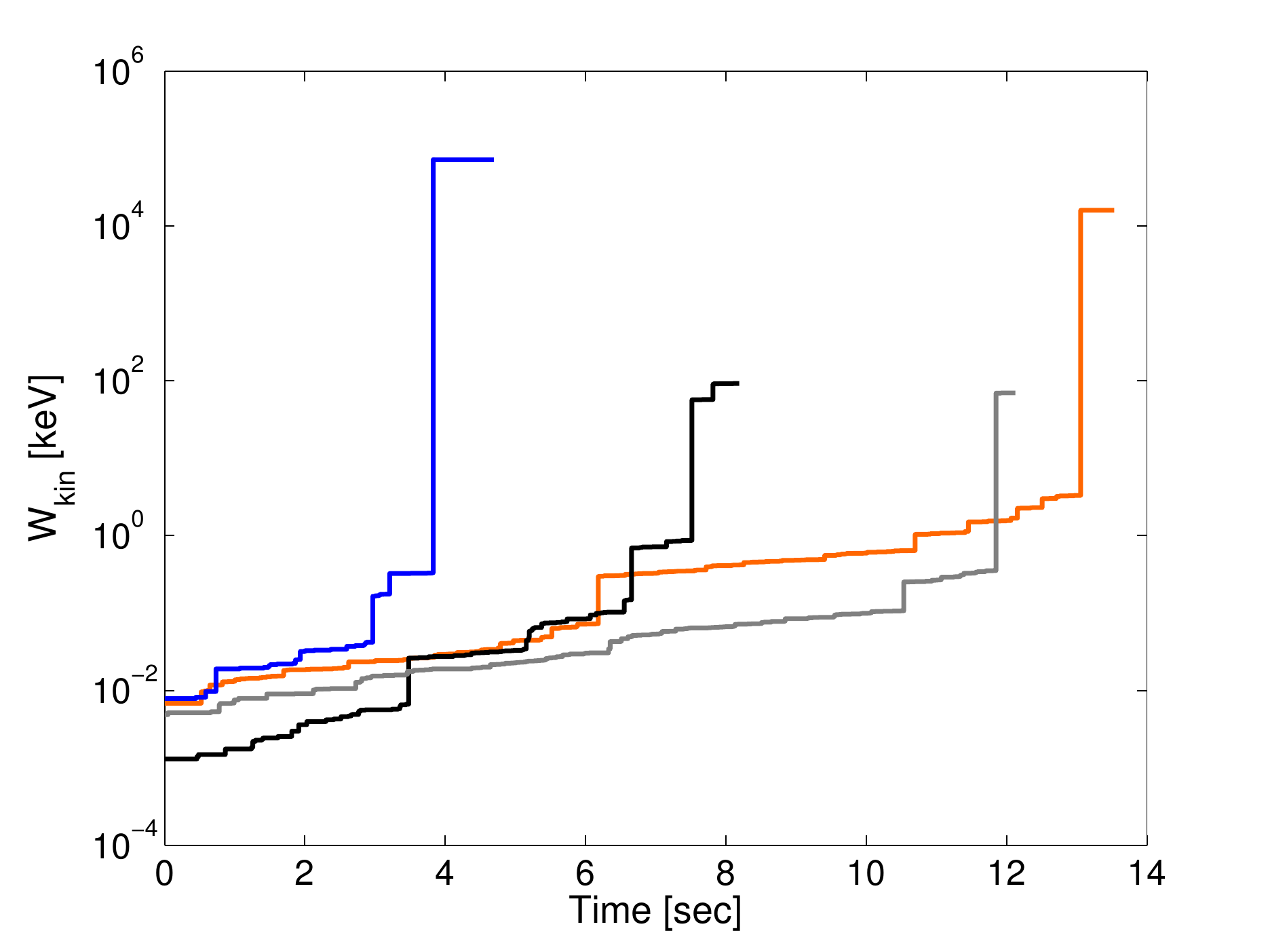}
\includegraphics[width=0.45\columnwidth]{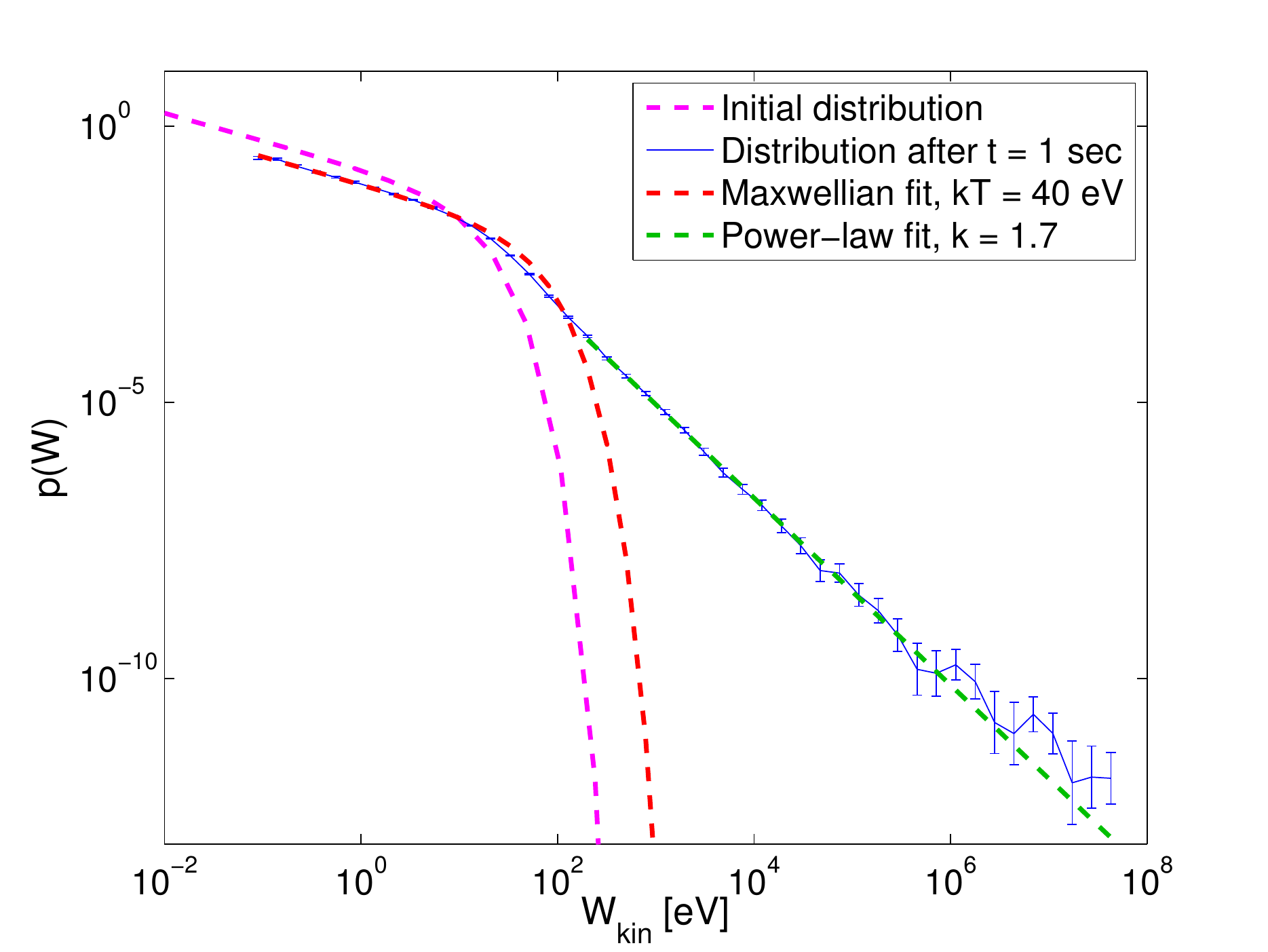}
\caption{ (a)  Energy as a function of time of a few selected particles. The particles gain energy systematically. (b) Energy distribution at $t=0$ and $t=1\,$s for the electrons staying inside the box. The initial temperature is $100$eV. \cite{Isliker17}.}
\label{Fermi3}
\end{figure}

Isliker et al.\ \cite{Isliker17} replaced the stochastic scatterers of Pisokas at al.\ with UCSs. The rest of the parameters and set-up are kept the same as in \cite{Pisokas16}. The UCSs are systematic (first order) scatterers, the particles gain energy when they interact with the UCSs. Isliker et al.\ approximate the {\bf macroscopic} energy gain as
\beq
\Delta W=| q |E_{eff}\ell_{eff} ,
\eeq
where $E_{eff} \approx (V/c) \delta B$ is a measure of the effective electric field of the UCS and $\delta B$ is the fluctuating magnetic field encountered by the particle, which is of stochastic nature, related to the stochastic fluctuations induced by reconnection. $\ell_{eff}$ is the characteristic length of the interaction of the particles with the UCS and should be proportional to $E_{eff}$, since small $E_{eff}$ will be related to small-scale UCS. 

Typical particle trajectories are shown in Fig.\ \ref{Fermi3}a, and the asymptotic energy distribution is shown in Fig.\  \ref{Fermi3}b.  The particles are  accelerated much faster  than in the stochastic acceleration process discussed above. The power law is now  softer and agrees well with the one from PIC simulations \cite{Guo15} or the results reported above from the tracking of particles inside the electromagnetic fields of 3D MHD codes \cite{Onofri06, Turkmani05, Arzner06}.

%todo: continue from here 
In turbulent reconnection, stochastic scattering at large scale disturbances co-exists with acceleration at UCSs. It is natural to ask how the ambient particles  react if the two Fermi accelerators act simultaneously. Pisokas et al.\  \cite{Pisokas18} discuss the synergy of energization at large scale magnetic disturbances  (stochastic scatterers) with the systematic acceleration by UCSs.  

\begin{figure}[h!]
\centering
\includegraphics[width=0.45\columnwidth]{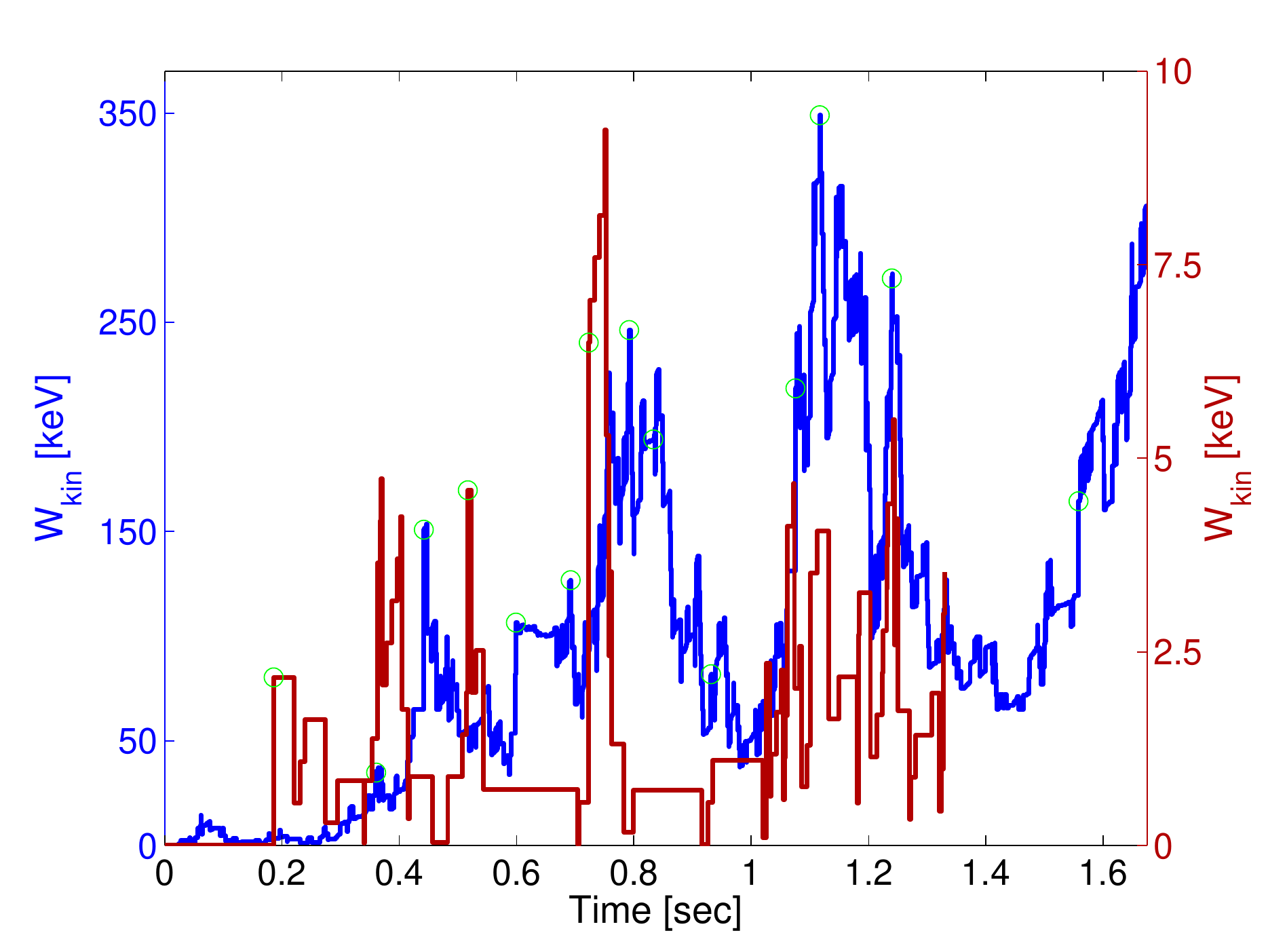}
\includegraphics[width=0.45\columnwidth]{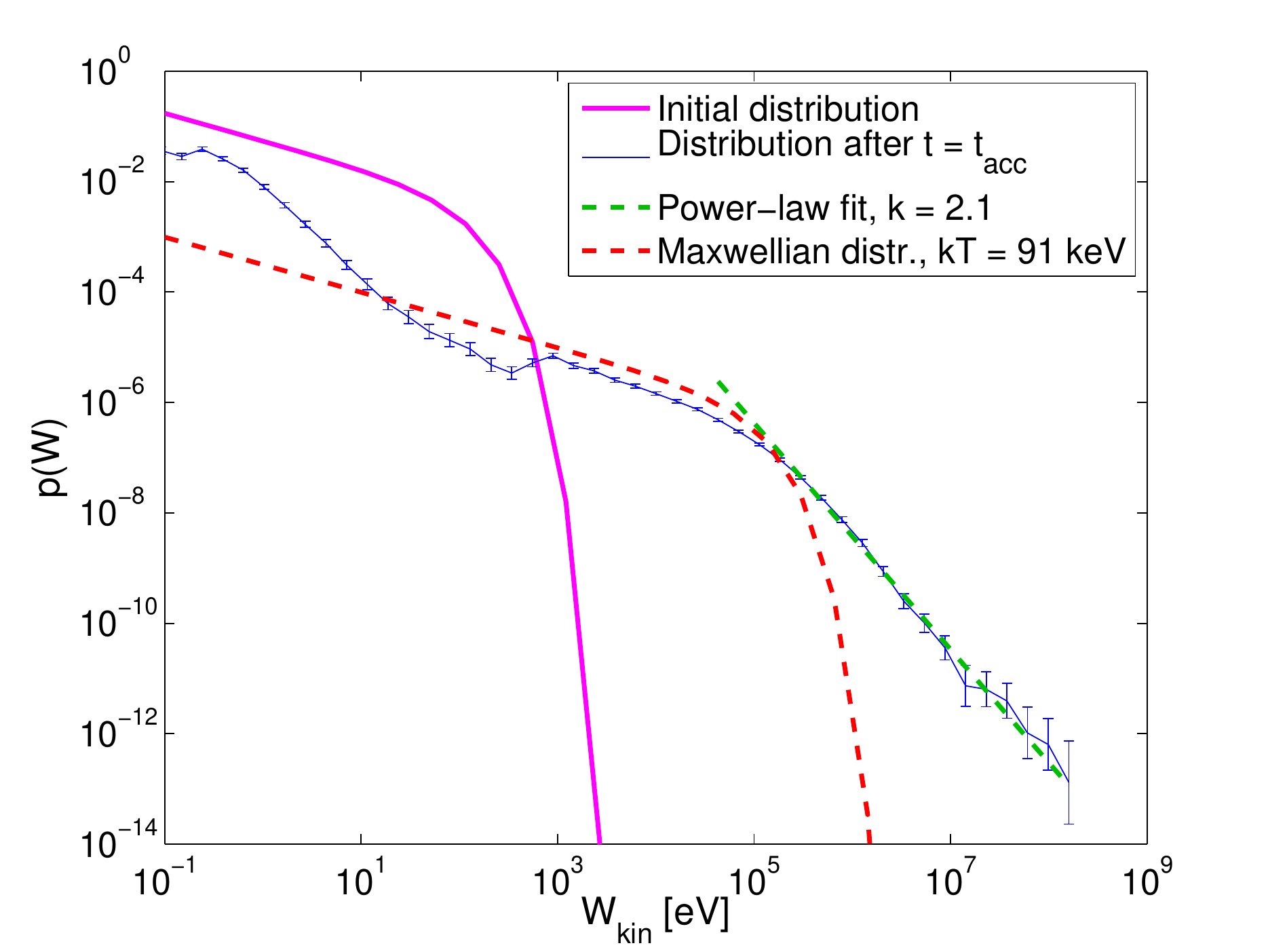}\\
\includegraphics[width=0.50\columnwidth]{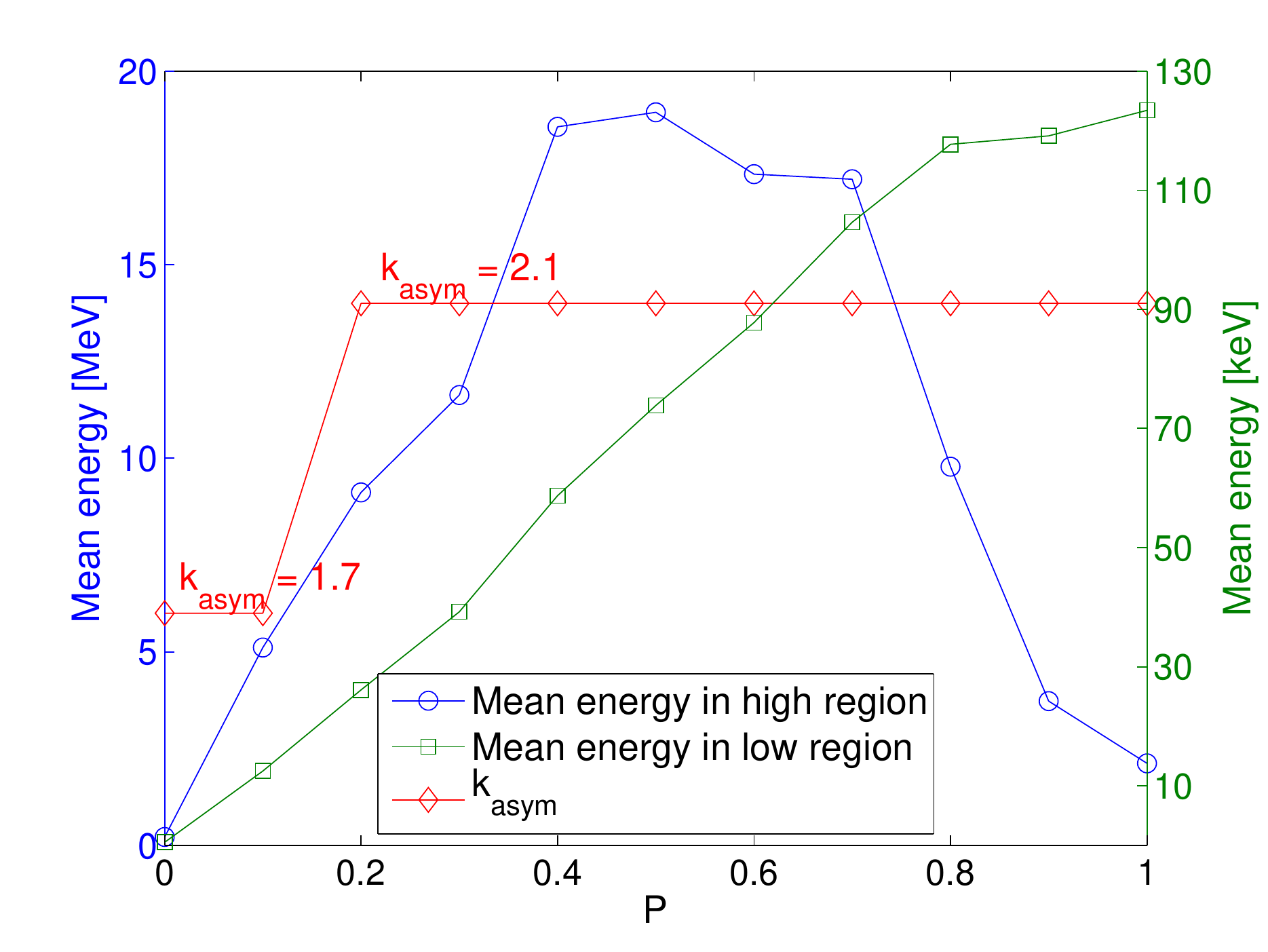}%\includegraphics[width=0.50\columnwidth]{Tfig3}
\caption{ (a)  Kinetic energy of typical particles as a function of time. (b) Energy distribution of the electrons that stay inside the box until $t=t_{esc} \sim 1.7 s$ (blue) for $P=0.5$, initial distribution (magenta), Maxwellian fit to the heated low - energy particles (red dashed), and fit  to the power law tail (green dashed). (c) Mean energy at $t=t_{esc}$ of the high energy tail (blue) and of the heated low energy particles (green), for different values of $P$ (fraction of the two kinds of scatterers). The red  points denote the asymptotic value $k_{asym}$ of the power-law index. \cite{Pisokas18}.}
\label{Fermi4}
\end{figure} 

They use the same modeling approach as in \cite{Pisokas16,Isliker17},
assuming that the charged particles scatter off the active grid points and gain or lose energy. The scatterers are divided into two classes. A fraction $P$ $(0 \leq P \leq 1 )$ are magnetic disturbances and the rest ($1-P$) are UCSs. When $P=0$ all scatterers are UCSs \cite{Isliker17} and the interaction with particles is systematic, and when $P=1$ all scatterers are magnetic disturbances and the interaction is stochastic \cite{Pisokas16}.   These two extreme cases have been discussed briefly  above. Pisokas et al.\  \cite{Pisokas18} use  a typical value $P=0.5$ and keep all the other parameters constant. 
%In Fig.\ \ref{Fermi4}a we follow the kinetic energy evolution of a few typical electrons. 
The synergy of the systematic and stochastic energization is obvious in the evolution of the kinetic energy of typical electrons, see Fig.\ \ref{Fermi4}a. In Fig.\ \ref{Fermi4}b  the energy distribution of the electrons is shown. The most striking characteristic is that heating and the formation of a high energy tail co-exist in the asymptotic distribution. Fig.\ \ref{Fermi4}c shows the power law index, the mean energy carried by the bulk of the hot plasma and by the tail as a function of $P$. The main result is that when $P$ is larger than 0.2 then the index of the power law tail remains constant and equals 2. This is a remarkable new result for astrophysical plasmas, since so far it was believed that only Diffusive Shock Acceleration can provide a stable power law index with this specific value. 
 The synergy of the two Fermi processes during turbulent reconnection  accelerates the electrons on sub-second time scales, the heating process though is much slower. It seems that heating is primarily due to the stochastic interaction of particles with magnetic disturbances and  the tail with its specific index is formed through a true synergy of the two mechanisms, possibly with a predominance, to some degree, of 
 	the systematic interaction of particles with UCSs.

%\subsection{Particle acceleration  by systematic Fermi processes}

\section{Summary}
In this review we have stressed the following points:

\begin{itemize}
	\item We present evidence from numerical simulations that supports the fact that all well known nonlinear structures (e.g.\ turbulence, current sheets, and shocks) asymptotically lead to a new nonlinear state, which we call \textit{turbulent reconnection}. 
	\item Turbulent reconnection is a non linear state of the plasma, where large scale magnetic disturbances and UCSs co-exist. 
	\item Based on numerical simulations that are still far from realistic, we suggest that the Solar convection zone may generate and drive the Solar corona into a turbulent reconnection state. The emergence of new magnetic flux, the random stressing of emerged magnetic flux by turbulent photospheric flows, and large scale instabilities of the emerged magnetic field topologies drive a variety of global and/or localized volumes into the state of  turbulent reconnection.
	\item The synergy of the large scale magnetic disturbances and the UCS in turbulent reconnection provides the heating and the acceleration of high energy particles (electrons and ions). We claim that during turbulent reconnection the two well known Fermi mechanisms (stochastic and systematic) co-exist, forming a new very efficient mechanism for the energization of the plasma.
	\item The stochastic interaction of the particles with the large amplitude magnetic fluctuations is responsible for the heating  and the synergy of stochastic and systematic acceleration for the formation of the high energy tail. 
		
	\item The key elements for the efficient heating and acceleration of particles are (1) the strength of the magnetic field in the energy release volume, (2) the mean free path $\lambda$ the particles travel between scatterers, (3) the size of the energy release volume.
\end{itemize}

The attempts made so far to analyse the solar corona using  simple  monolithic magnetic topologies, i.e.\ a single loop, a single current sheet or a shock, fail to grasp the importance of the turbulent state of the solar corona and its consequences, as discussed here. We hope that the Parker Solar Probe will capture the dynamics of the fully developed turbulence in the Solar Corona and let us discover the way it is coupled to the turbulent solar wind.

The results reported here can be applied to many astrophysical, space or laboratory plasmas, whenever the state of turbulent reconnection is established.

\section{Acknowledgments}	This work was supported by the national Programme for the Controlled Thermonuclear Fusion, Hellenic Republic. The sponsors do not bear any responsibility for the content of this work.

	\section{Referecnes}%\nocite{*}
	%\bibliographystyle{iopart-num}
%\bibliography{vlahosastro}% Produces the bibliography via BibTeX.
\providecommand{\newblock}{}

\end{document}